\documentclass[twocolumn]{aastex63}
\usepackage{amsmath}
\usepackage{amssymb}
\usepackage{bm}
\usepackage{graphicx}
\usepackage{xcolor}

\shorttitle{Modeling Weak Lensing with Imaging and Kinematics}
\shortauthors{DiGiorgio et al.}

\begin{document}

\title{A Novel Framework for Modeling Weakly Lensing Shear Using Kinematics and Imaging at Moderate Redshift}

\correspondingauthor{Brian DiGiorgio}
\email{bdigiorg@ucsc.edu}

\author[0000-0003-3344-7776]{Brian DiGiorgio}
\affil{Department of Astronomy and Astrophysics, University of California, 1156 High Street, Santa Cruz, CA 95064, USA}

\author{Kevin Bundy}
\affiliation{Department of Astronomy and Astrophysics, University of California, 1156 High Street, Santa Cruz, CA 95064, USA}
\affiliation{University of California Observatories, University of California, 1156 High Street, Santa Cruz, CA 95064, USA}

\author{Kyle B. Westfall}
\affiliation{University of California Observatories, University of California, 1156 High Street, Santa Cruz, CA 95064, USA}

\author{Alexie Leauthaud}
\affil{Department of Astronomy and Astrophysics, University of California, 1156 High Street, Santa Cruz, CA 95064, USA}

\author{David Stark}
\affil{Department of Physics and Astronomy, Haverford College, 370 Lancaster Avenue, Haverford, PA 19041, USA}

\begin{abstract}

Kinematic weak lensing describes the distortion of a galaxy's projected velocity field due to lensing shear, an effect recently reported for the first time by Gurri et al. based on a sample of 18 galaxies at $z \sim 0.1$. In this paper, we develop a new formalism that combines the shape information from imaging surveys with the kinematic information from resolved spectroscopy to better constrain the lensing distortion of source galaxies and to potentially address systematic errors that affect conventional weak-lensing analyses. Using a Bayesian forward model applied to mock galaxy observations, we model distortions in the source galaxy's velocity field simultaneously with the apparent shear-induced offset between the kinematic and photometric major axes. We show that this combination dramatically reduces the statistical uncertainty on the inferred shear, yielding statistical error gains of a factor of 2--6 compared to kinematics alone.  While we have not accounted for errors from intrinsic kinematic irregularities, our approach opens kinematic lensing studies to higher redshifts where resolved spectroscopy is more challenging.  For example, we show that ground-based integral-field spectroscopy of background galaxies at $z \sim 0.7$ can deliver gravitational shear measurements with S/N $\sim 1$ \textit{per source galaxy} at 1 arcminute separations from a galaxy cluster at $z \sim 0.3$. This suggests that even modest samples observed with existing instruments could deliver improved galaxy cluster mass measurements and well-sampled probes of their halo mass profiles to large radii.

\end{abstract}


\section{Introduction} \label{sec:intro}
Weak gravitational lensing is a powerful and increasingly utilized tool for measuring how mass is distributed throughout the Universe \citep[e.g.][]{mandelbaum18}. Applications include use of cosmic shear to probe the mass density and the growth of structure in the Universe (e.g. \citealt{desy1, desy3, hikage19}), galaxy-galaxy lensing to characterize halo masses of galaxies in different stellar-mass regimes and perform $3 \times 2$ correlation studies (e.g. \citealt{alexie12, krause17}), and cluster lensing profiles to measure cluster-scale halo mass profiles and halo shapes \citep[e.g.][]{umetsu18, mandelbaum06}. These science goals have motivated large-scale weak-lensing surveys with new instruments, telescopes, and even satellite missions.  Major projects include the Hyper Suprime Cam \citep[HSC,][]{aihara18}, the Kilo Degree Survey \citep[KiDS,][]{kuijken15}, the Legacy Survey of Space and Time \citep[LSST,][]{lsst}, \textit{Euclid} \citep{laureijs11}, the Nancy Grace Roman Space Telescope \citep{wfirst}, and the Dark Energy Survey \citep[DES,][]{krause17, desy1, desy3}. 

Weak-lensing surveys seek to measure distortions in a galaxy's shape caused by lensing shear, a geometric effect caused by the gravitational lensing of a background source by mass in the foreground. To first order, this distortion results in background galaxies becoming elongated in the direction tangential to the surface-density gradient of the foreground mass distribution. 

A number of factors complicate lensing measurements in conventional surveys that use imaging to measure galaxy shapes.  Weak lensing distortions only induce a ${<}1\%$ change in observed ellipticity.  Measurements at this level of precision are typically achieved by statistically stacking results from multiple sources, but our uncertain knowledge of the intrinsic or ``pre-lensed'' galaxy ellipticity distribution (``shape noise'') typically necessitates thousands of stacked sources to detect a lensing signal.  
Potential biases become a greater concern with added complications, such as intrinsic alignments of source galaxies with near-foreground dark matter structure (e.g., filaments) and systematic errors in shape measurements resulting from instrumentation \citep[e.g.,][]{troxel15,mandelbaum18}. 
Very large samples are therefore required to drive down statistical errors and test for systematics that can otherwise overwhelm the shear signal.

For instance, the CLASH survey \citep{umetsu14} used ${\sim}10^4$ background galaxies around each of 20 foreground Abell clusters to recover mass density profiles precise enough to constrain dark matter halo mass profiles. Meanwhile, cosmic shear surveys like the DES \citep{desy1} use ${\sim}10^7$ galaxies to sufficiently characterize their signal, but even these large samples remain susceptible to systematic biases. All of these efforts require deep imaging over wide areas, typically carried out on 2 meter- to 8 meter-class telescopes.g

The burgeoning subfield of kinematic weak lensing (KWL), also referred to as ``precision weak lensing'' in the literature \citep{gurri20, gurri21}, provides an additional means of inferring lensing shear by measuring distortions in the projected velocity field of source galaxies.  Although KWL requires more expensive spectroscopic observations, many fewer galaxies are needed to detect a signal.  This is because the velocity field encodes a trace of the galaxy's original, pre-lensed coordinate space.  The induced lensing distortions can be fit directly, eliminating or at least dramatically reducing the shape noise.  If future weak-lensing analyses could make use of KWL, they could enable higher spatial resolution maps of foreground mass, more sensitive mass measurements, and independent checks on results from conventional imaging surveys \citep{huff20}.

The literature investigating applications of KWL has fallen into two regimes: high-spatial-resolution kinematic measurements on small samples, and lower-precision kinematic measurements derived from large samples. The idea of KWL was first put forward by \citet{blain02}, who determined that weak lensing shear would change the symmetry of a galaxy's rotational velocity as measured in an azimuthal ring of constant radius. The idea was further developed by \citet{morales06}, who proposed measuring the angle between the kinematic major and minor axes, which are no longer perpendicular in a lensed galaxy. A similar idea was developed by \citet{dbd15}, who suggested searching for lensing-induced asymmetries in the reflection symmetries of galaxy kinematic data. All these authors emphasize the potential of making a shear measurement with a single galaxy and obtaining a measurement that would be independent of shape noise. However, none of these studies reported a detection due to the small size of lensing effects at nearby redshifts where the high-spatial-resolution spectroscopic measurements required for precise measurements are most readily available.

\citet{huff20} instead proposes implementing KWL on a survey scale, considering large samples where individual measurements with less kinematic information are statistically stacked.
Building on the work of \cite{huff13}, they propose targeted measurements of lensing-induced differences in the projected velocity along the major and minor axes.
\cite{wittman21} performs a Fisher Matrix analysis of this technique in a hypothetical DES-scale survey to derive theoretical limits on the covariances of the derived lensing parameters. 

In the first reported detection of kinematic weak lensing, \cite{gurri20} stake out a middle ground in sample size versus per-galaxy information content.
They collect and analyze 2D velocity fields of ${\sim} 20$ galaxies at $z < 0.15$, and report a positive mean shear amplitude detected at $2.5\sigma$. They forward-model each source galaxy with a rotating thin-disk model that allows shear to vary and apply their technique to selected source galaxies likely to be sheared by foreground halos.  The detection reported by  \cite{gurri20} is an exciting development in the young field of KWL, but also highlights upcoming observational challenges. For instance, they find  discrepancies between their theoretical and observed shear magnitudes, which they say likely arise from a combination of intrinsic kinematic irregularities in their sample galaxies (``dynamical shape noise") and the scatter in the stellar-halo mass relation \citep[see][]{gurri21}. It will be difficult to obtain the kinds of highly-sampled and high-S/N velocity fields in \cite{gurri20} for source galaxies at $z \lesssim 1$, but it is at these redshifts where probes of structure formation are most needed and the lensing kernel more favorable \citep{weinberg13}.


With this challenge in mind, we build on the techniques of \cite{gurri20} by introducing additional constraints from galaxy shape measurements into the KWL formalism using a Bayesian forward model.
We show that by including measurements of the major-axis position angle of an ellipse fit to the surface brightness profile and comparing it to the position angle derived for the sheared velocity field, we can improve the per-galaxy shear S/N by several times in many cases.  This motivates us to explore the potential use case of measuring the mass and shape of galaxy cluster halos at $z \sim 0.3$ \citep{bartelmann17} with IFU observations of 50--100 galaxies at $z \sim 0.7$.  Immediately valuable for studies of individual clusters, a future survey program could aid mass calibrations required for cluster cosmology \citep[e.g.][]{sptsz19} and opens new possibilities for wide-scale KWL surveys at $z \lesssim 1$ with a dedicated survey.



The paper is structured as follows: Section \ref{sec:conventional} gives background on relevant weak-lensing theory and introduces our formalism for how lensing affects the shape of a galaxy. Section \ref{sec:kwl} studies the impact of lensing shear on the shape of a rotating galaxy and compares the merits of different techniques for measuring kinematic shear observables. Section \ref{sec:phot} develops our modeling framework for including offsets between the kinematic axes and axes derived from imaging for a given galaxy and characterizes the improvement of the precision of lensing measurements and the important increase of S/N that results. Section \ref{sec:conclusions} gives a summary and looks forward to future applications of the technique.

Throughout this paper, we assume a Planck 2018 cosmology \citep{planck18} as implemented in the \texttt{Colossus} cosmology package from \citet{diemer18}, with $H_0 = 67.36$ km s$^{-1}$ Mpc$^{-1}$ and $\Omega_m = 0.3111$ at $z=0$.

\section{Weak Lensing Effects on Imaging} \label{sec:conventional}

To understand the benefits image position angle measurements can have for KWL, we first review key aspects of how lensing geometry changes the shape of galaxies. Imaging-based lensing surveys typically model background galaxies as ellipses with a measured on-sky position angle and ellipticity that are used to derive lensing amplitude. This serves both as context for the current state of the field and as a basis for the techniques developed in Section \ref{sec:phot}. For a more detailed treatment of the subject, see \cite{bartelmann17}.

\subsection{General Lensing Theory}
Gravitational lensing is a well-established result of general relativity and its effects have been well-characterized \citep{me91}. Because photons travel along geodesics, when they pass through regions of spacetime that have been distorted by a mass distribution, the geodesic is deflected, curving their paths. So the observed image of a background galaxy will be distorted by any foreground mass along the line of sight.

If we assume that the characteristic scale of the foreground mass distribution is small compared to the other relevant distance scales in the system --- the angular diameter distance from the observer to the foreground mass, $D_L$, the distance from the observer to the background source, $D_S$, and the distance from the foreground mass to the background source, $D_{LS}$ --- we may adequately describe the system using the thin lens approximation. This stipulates that the deflection of the photons in the system happens in an infinitely thin plane located at the lens position and that the photon travels in a straight line everywhere else. The magnitude of the angular deflection relates to the projected surface density $\Sigma$ of the 3D foreground mass distribution $\rho$, where 

\begin{equation}
	\Sigma(\bm{\xi}) = \int\rho(\bm{\xi, z}) d\bm{z},
\end{equation}

\noindent for $\bm{\xi} = \sqrt{\bm{x}^2+\bm{y}^2}$ with $\bm{x}$ and $\bm{y}$ being the physical coordinates of the mass distribution on-sky and $\bm{z}$ being the line of sight vector from the observer. The angular deflection of a photon is given by 

\begin{equation}
	\bm{\alpha}(\bm{\xi}) = \frac{4G}{c^2}\int \Sigma(\bm{\xi}') \frac{\bm{\xi-\xi'}}{|\bm{\xi-\xi'}|^2}\, d^2\bm{\xi}'.
\end{equation}

\noindent The deflection angle can also be given in terms of the lensing potential $\psi$ as 

\begin{equation}
	\bm{\alpha}(\theta) = \frac{D_S}{D_{LS}} \nabla_\theta \psi,
\end{equation}

\noindent where $\theta$ is the angular separation of the source as seen from the perspective of the observer: $\theta = |\bm{\xi}|/D_L$. 
For a point mass lens in the weak lensing regime,
$\psi$ can be written as

\begin{equation} \label{psi}
	\psi = \frac{4GM}{c^2}\frac{D_{LS}}{D_L D_S} \ln |\bm{\xi}|.
\end{equation}

The magnitude of the lensing effect is often expressed in terms of the critical surface mass density, which defines the characteristic angular scale of a lensing system:  

\begin{equation} \label{sigcrit}
	\Sigma_\textrm{crit} = \frac{c}{4\pi G}\frac{D_S}{D_L D_{LS}}.
\end{equation}

\noindent The distortion manifests itself in two ways. The first is the convergence $\kappa$, which has magnitude $\kappa = \Sigma/\Sigma_\textrm{crit}$ and magnifies the background source. However, for the purposes of this paper, we will be more interested in the lensing shear.

The shear, as its name implies, distorts the coordinate system of the source, changing its observed shape. This distortion can be broken into two separate components defined by their relation to the derivatives of the lensing potential: 

\begin{equation}
\gamma_+ = \frac{1}{2}\left(\frac{\partial^2\psi}{\partial\theta_x^2} - \frac{\partial^2\psi}{\partial\theta_y^2}\right), \ \ \ \gamma_{\times} = \frac{\partial^2\psi}{\partial\theta_x \partial\theta_y}.
\end{equation}

\noindent where $\theta_x$ and $\theta_y$ are perpendicular angular coordinates in the arbitrarily defined coordinate system projected onto the lens plane. Following the formalism of \citet{huff13}, the shear aligned with the coordinate axes is called $\gamma_+$, while the cross term $\gamma_{\times}$ is aligned with the diagonals of the coordinate system. For a simplified case of an axisymmetric lens, the shear magnitude is given by

\begin{equation} \label{shear}
|\gamma| = \frac{\Delta \Sigma}{\Sigma_\textrm{crit}},
\end{equation}

\noindent where $\Delta \Sigma$ is the differential projected surface mass density, the difference between the value of $\Sigma$ at a given radius $r_0$ and the mean $\Sigma$ within that radius:

\begin{equation}
	\Delta \Sigma = \bar{\Sigma}(r < r_0) - \Sigma(r_0).
\end{equation}

\noindent In the simplified axisymmetric case, the two shear components can also be expressed as

\begin{equation} \label{sheardefs}
\gamma_+ = |\gamma|\cos 2\theta\ \ \ \gamma_\times = |\gamma|\sin 2\theta,
\end{equation}

\noindent with $\theta = \tan^{-1} (\theta_y/\theta_x)$. Positive and negative values for the shear result in distortions in opposite directions. We assume $|\gamma| < 0.1$ in this paper to remain in the weak-lensing regime where the magnitude of the distortions are small and the small angle approximation is still valid.

\subsection{Image Distortions} \label{sec:phottheory}

\begin{figure}
	\centering
	\includegraphics[width=\linewidth]{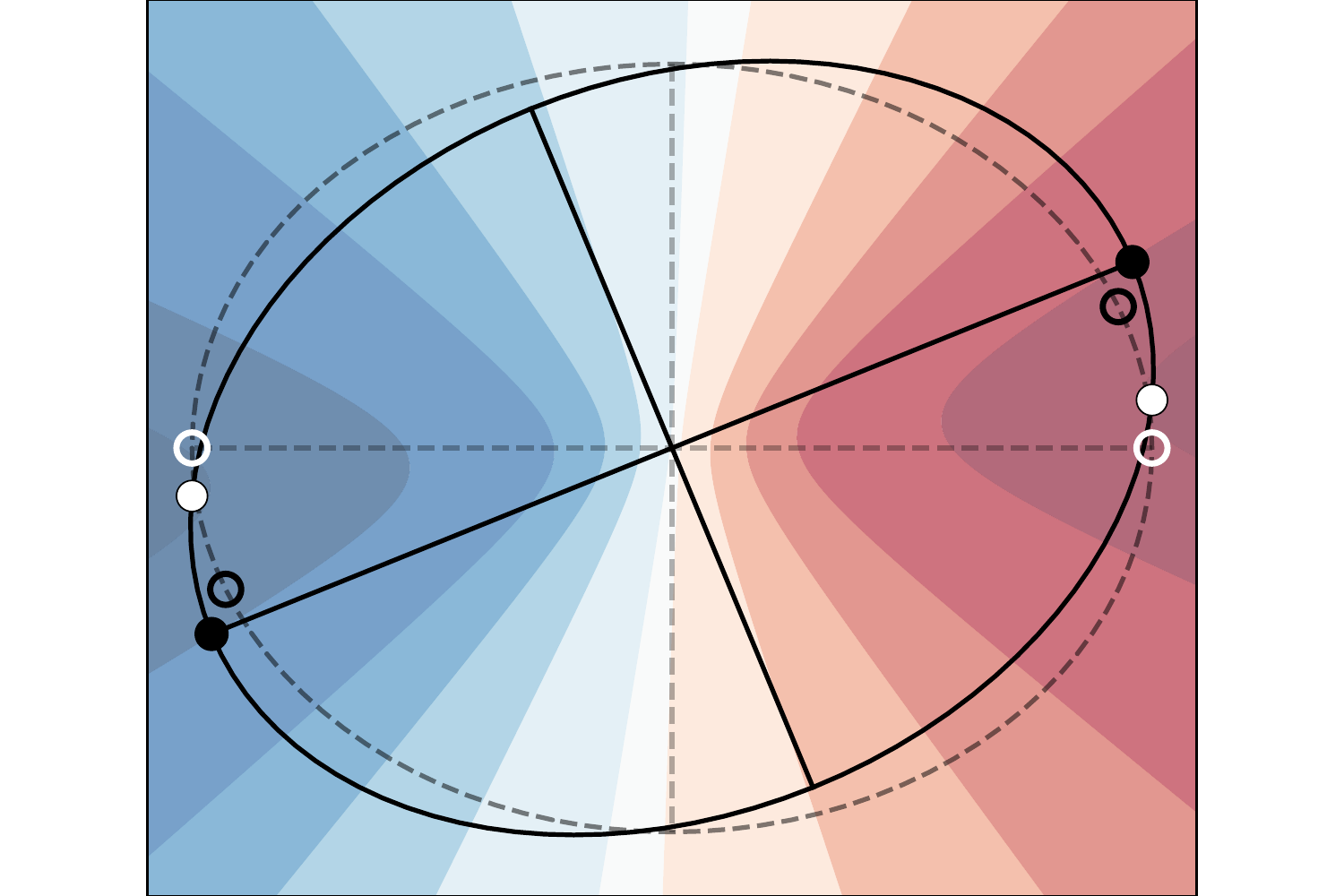}
	\caption{Representation of how shear from gravitational lensing affects the positions of major and minor axes derived from kinematics and imaging/photometry. The original orientation of the galaxy is shown as a dashed elliptical isophote, with the dashed lines marking the positions of the unlensed major and minor axes (photometric and kinematic are the same). A lens to the upper left of the galaxy applies a shear, distorting the galaxy's intrinsic coordinate system and velocity field contours according to the transformation matrix $A$ (Equation \ref{Amatrix}). The solid ellipse represents how a new elliptical isophote would be fit to the sheared galaxy, along with its new imaged major and minor axes. The colored contours represent the observed velocity field, with red and blue indicating receding and approaching sides of the galaxy, respectively, and darker colors indicating higher velocity magnitude. The point along the lensed isophote that is measured as the lensed imaged major axis (filled black circle) was not on the major axis of the unlensed galaxy (unfilled black circle), leading to a discrepancy between the lensed axis observed in imaging and the kinematic axis (filled white circle). A similar effect happens with the minor axis. The galaxy shown has an axis ratio $q = 0.8$ and the shear applied is $\gamma_\times = 0.1$.}
	\label{example}
\end{figure}

In the weak-lensing regime, the recognizable features of strong lensing, like arcs and multiple images, give way to tiny distortions in background sources that are usually only detectable through statistical methods applied to large samples. Weak lensing effects are well-parameterized by a single transformation matrix that can be applied to the coordinate system of the background source:

\begin{equation} \label{originalAmatrix}
A=
\begin{pmatrix}
	1-\kappa+\gamma_+ & \gamma_\times \\
	\gamma_\times & 1-\kappa-\gamma_+
\end{pmatrix}
\end{equation}

\noindent If we do not care about the size and luminosity of the source, we may disregard the convergence $\kappa$ since it is small in the weak-lensing regime, so we may simplify $A$ to be: 

\begin{equation}\label{Amatrix}
A\approx
\begin{pmatrix}
1+\gamma_+ & \gamma_\times \\
\gamma_\times & 1-\gamma_+
\end{pmatrix}.
\end{equation}

\noindent The effect of $\gamma_+$ is to symmetrically elongate the source along the tangential axis\footnote{Note that this is distinct from $\kappa$, which elongates the galaxy in all directions equally, causing a perceived magnification.}, meaning the on-sky orientation of the source is not substantially affected. However, $\gamma_\times$ shears the source, causing an effective rotation.

We can apply this transformation to a toy model of a galaxy, a typical background source in a weak-lensing survey. We may model the galaxy as a circle inclined at some angle relative to the observer, so it is seen simply as a series of concentric elliptical isophotes, each satisfying

\begin{equation}
	1 = q^2x^2 + y^2,
\end{equation}

\noindent where $q$ is the axis ratio of the projected ellipse. Applying the shear transformation matrix to this object, we find that, post-lensing, the coordinate system has become $\bm{x}' = A^{-1}\bm{x}$ for $\bm{x} = (x,y)^T$, so 

\begin{equation} \label{transform}
	1 \approx q^2x'^2(1-2\gamma_+) + y'^2(1+2\gamma_\times) - 2x'y' \gamma_\times (1+q^2),
\end{equation} 

\noindent where we have dropped higher order terms in $\gamma_+$ and $\gamma_\times$ since we expect the magnitude of these terms to be small in the weak lensing regime.

We then treat the deformation of the elliptical isophote as a rotation rather than a shear. The general equation of an ellipse rotated by angle $\alpha$ relative to the origin is

\begin{multline}
	1=\left(\frac{\cos^2\alpha}{a^2} + \frac{\sin^2\alpha}{b^2}\right)x^2 + \left(\frac{\sin^2\alpha}{a^2} + \frac{\cos^2\alpha}{b^2}\right)y^2 \\
	 - 2 \cos\alpha \sin\alpha \left(\frac{1}{b^2} - \frac{1}{a^2}\right), 
\end{multline}

\noindent where $a$ and $b$ are the lengths of the major and minor axes, respectively. We may greatly simplify this equation if we restrict ourselves to the weak-lensing regime. By assuming $\alpha$ to be small, we can apply the small angle approximation and ignore higher order terms in $\alpha$. If we then multiply through by $b^2$ to put everything in terms of the axis ratio $q = b/a$ and set $b=1$ in arbitrary units, we have 

\begin{equation} \label{rotate}
	1 \approx q^2 x^2 + y^2 - 2\alpha xy\left(1-q^2\right).
\end{equation}

Equations \ref{transform} and \ref{rotate} are identical except for the cross term. If we assume that the shear can indeed be thought of as a rotation, we can set these two equations equal to each other to find 

\begin{equation}\label{angphot}
	\alpha (1-q^2) \approx \gamma_\times (1+q^2)\ \ \ \Rightarrow\ \ \ \alpha \approx \gamma_\times \frac{1+q^2}{1-q^2}.
\end{equation}

\noindent So the effect of the shear on an isophote can be well-approximated in the weak-lensing regime as a rotation, as seen in Figure \ref{example}. The rotation angle is determined by $\gamma_{\times}$, which depends on the magnitude of the overall shear, the relative positions of the foreground lens and the background source, and the axis ratio of the background source, which are in turn determined by Equation \ref{sheardefs} and the inclination of the galaxy. The magnitude of $\alpha$ is maximized when the background source has a low inclination, has a position angle that is misaligned with the radial vector by $45^\circ$, and when the foreground lens is more massive.

The axis ratio $q$ changes as a result of lensing as well, as seen by the effect of the transformation matrix $A$ on a point at $(a,b)$:

\begin{multline}
	\begin{pmatrix}
	1+\gamma_+ & \gamma_\times \\
	\gamma_\times & 1-\gamma_+
	\end{pmatrix}
	\begin{pmatrix}
	a \\ b
	\end{pmatrix}
	\\ = 
	\begin{pmatrix}
	(1 + \gamma_{\times} + \gamma_+)\, a \\
	(1 + \gamma_{\times} - \gamma_+)\, b
	\end{pmatrix}
	=
	\begin{pmatrix}
	a' \\ b'
	\end{pmatrix},
\end{multline}

\noindent where we have defined $(a', b')$ as the major and minor axis lengths of the lensed galaxy. Thus, the lensed axis ratio $q'$ is

\begin{equation}
	q' = \frac{b'}{a'} = \frac{(1 + \gamma_{\times} - \gamma_+)}{(1 + \gamma_{\times} + \gamma_+)} \frac{b}{a}  \\
	\approx \left(1- \frac{2 \gamma_+}{1 + \gamma_{\times}}\right) q,
\end{equation}

\noindent where we have again exploited a Taylor series expansion to make a first order approximation. What observers see is the post-lensing axis ratio $q'$ rather than $q$, so we again approximate to write this as

\begin{equation} \label{qdistortion}
	q \approx \frac{q'}{1-2\gamma_+}.
\end{equation}

\noindent We see that for small $\gamma_+$, $q \approx q'$, and because we will be concerned mostly with cases where this is true, we will ignore the distinction between pre- and post-lensing axis ratios for the remainder of the paper.

However, neither the angular distortion nor the axis ratio distortion can be directly measured in images of any single galaxy because nothing is known about its original shape. Before the galaxy is lensed, it has some intrinsic position angle and ellipticity that provide the baseline for any lensing distortions. With conventional imaging surveys, the only way we can perceive changes is by looking at the statistical properties of the sample in aggregate. A large sample of galaxies under the same shear will show a net alignment perpendicular to the induced shear that deviates from a random distribution of position angles. Measuring this net alignment is a primary goal of imaging-based weak lensing surveys. 

\section{Weak Lensing Effects on Kinematics} \label{sec:kwl}

While it is impossible to tell the difference between an elliptical isophote that has been rotated and elongated due to lensing shear and one that was never sheared, the same is not true for projected kinematics of a rotating disk. By taking spectroscopic measurements across the face of a rotating galaxy, the relative velocity of different parts of the galaxy can be determined from the Doppler shift of spectral features compared to the systemic velocity of the galaxy.\footnote{In this paper, we assume an infinitely thin, rotating disk. This assumption may cause problems with highly inclined disks.} The kinematic measurements that make up the velocity field are associated with specific coordinates in the galaxy's intrinsic plane, a relationship that is not broken by gravitational lensing. This probe of the galaxy's intrinsic geometry enables much more precise per-galaxy lensing measurements.


In what follows, our analysis is motivated by reducing the final statistical errors on a galaxy's shear measurement, as limited by the quality of the data.  We do not treat here the per-galaxy error that stems from intrinsic irregularities in the physical structure of the galaxy. Features that deviate from the model used in the fit will bias the results regardless of the quality of the data.  However, this dynamical shape noise \citep{gurri20} is expected to be randomly distributed across a sample and should beat down as the sample size increases.  We will return to the role of statistical and systematic errors in Section \ref{sec:gurri}.

\subsection{Kinematic Axis Distortion} \label{sec:radon}

Unlike the photometric measurements made by fitting isophotes, the transformation of a galaxy's velocity field cannot be modeled simply as a rotation of the projected field. In order to illustrate the effect of shear on the velocity field, we examine the behavior of two reference locations: a point on the major axis and one on the minor axis.

A point on the major axis of the unlensed galaxy (unfilled white circle on Figure \ref{example}) can be described by the coordinates $(x,0)$ for a rectilinear coordinate system aligned with the major and minor axes. We define the  on-sky angle of the major axis in a particular reference frame to be the kinematic position angle (PA). Applying the transformation matrix $A$ to this coordinate location, we find that this point gets moved to ${((1+\gamma_+)\,x,\ \gamma_\times x)}$ (filled white circle), so the angular displacement of the lensed major axis from the original major axis is 

\begin{equation} \label{majkin}
	\theta_{maj} = \tan^{-1} \left(\frac{\gamma_\times}{1+\gamma_+}\right) \approx \frac{\gamma_\times}{1+\gamma_+},
\end{equation}

\noindent where we have performed a Taylor series expansion of the inverse tangent and taken just the first order since we expect the shear to be small in the weak-lensing regime. 

We can do a similar transformation to a point on the unlensed minor axis $(0,y)$ to find 

\begin{equation} \label{minkin}
	\theta_{min} = \tan^{-1}\left(\frac{1-\gamma_+}{\gamma_\times}\right) \approx \frac{\pi}{2} - \frac{\gamma_\times}{1-\gamma_+},
\end{equation}

\noindent where we have again performed a first-order Taylor series expansion.

We can again exploit Taylor series to expand Equations \ref{majkin} and \ref{minkin} with respect to $\gamma_+$:

\begin{equation} \label{angkin}
	\theta_{maj} \approx \gamma_{\times}(1 - \gamma_+);\ \ \ \ \ \theta_{min} \approx \frac{\pi}{2} - \gamma_{\times}(1 + \gamma_+).
\end{equation} 

\noindent The dependence of the angular differences on $\gamma_+$ is on the order of $|\gamma|^2$ (assuming that $\gamma_+ \approx \gamma_{\times}$, which is true for most galaxy-lens orientations). This means that we can neglect the $\gamma_+$ term\footnote{The same is true for the $\kappa$ term that we have already neglected. If we had kept it, it would have shown up in the denominator here as well and could be neglected with the same logic.} and just write

\begin{equation} \label{angkinsimp}
	\theta_{maj} \approx \gamma_{\times}; \ \ \ \ \ \theta_{min} \approx \frac{\pi}{2} - \gamma_{\times}.
\end{equation}

\noindent These simple relations allow us to estimate the shear by measuring how much the axes differ from being orthogonal, as was proposed by \cite{morales06}. In an ideal velocity field, the angle between the major and minor kinematic axes should be 

\begin{equation} \label{perp}
	\theta_\perp \approx \frac{\pi}{2} \pm 2\gamma_{\times}.
\end{equation}

\noindent If we could determine the position angles of the velocity maximum, minimum, and zeros and see how they differ from perpendicularity using a technique like the Radon transform \citep{stark18} or kinemetry \citep{kinemetry}, we could use the above result to gain an estimate of the induced shear. 

\subsection{Sheared Velocity Field Fitting}  \label{sec:data}

We can produce a more complete picture of the shear of a velocity field if we model the distortion of the entire galaxy rather than just its major and minor axes. We construct a Bayesian forward model that allows us to utilize all of the velocity measurements to obtain a best-fit sheared velocity field.

For the purpose of comparing different KWL techniques in our regime of interest, we create a simulated observation of a galaxy. We consider a  source galaxy at $z = 0.7$ behind a $7 \times 10^{14}\ M_\odot$ foreground cluster (approximately equal in mass to the Coma cluster, \citealt{coma}) at $z \sim 0.3$. We assume the mass profile of the cluster follows a Navarro-Frenk-White profile \citep[NFW,][]{nfw97}. Two-dimensional emission-line kinematics are feasible at this redshift with 8m-class ground-based optical telescopes (e.g., \citealt{contini16}). We assume an impact parameter of 0.3 Mpc in the plane of the lens corresponding to a 1 arcminute separation from the cluster center, which is the field of view in the MUSE wide-field configuration \citep{muse}. This system yields a shear magnitude of $|\gamma| = 0.0589$. We assume that the kinematic major axis of the background galaxy is at a 45$^\circ$ angle to the radial vector to the lens, meaning $\gamma_{\times} = 0.0589$ and $\gamma_+ = 0$. With this shear magnitude in mind, we explore what types of measurements are possible.

\begin{figure}
	\centering
	\includegraphics[width=.9\linewidth]{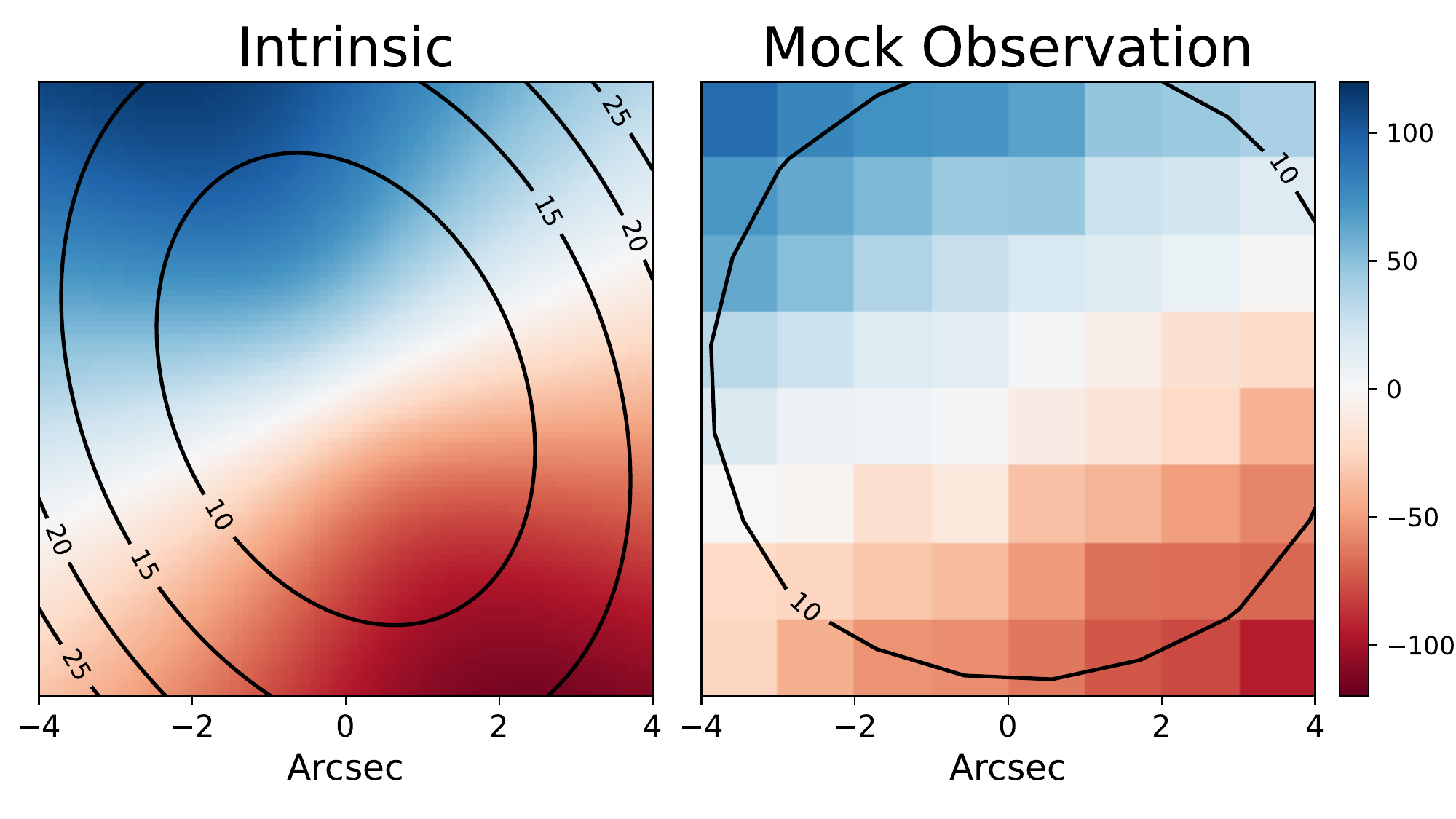}	
	\caption{
	Left: A mock intrinsic velocity field to test our  KWL shear extraction. This galaxy was generated using the model in Equation \ref{vfmodel} with parameters $v_{max} = 220$ km/s , $i = 45^\circ$, $\phi=2$ rad, $h=2$ kpc, and $v_{sys}=0$ km/s. A shear of $\gamma_{\times}=0.0589$ has been applied to it. The overlaid contours show the error on the velocity measurements in km/s, which are normalized to a central value of 5 km/s. The inverse variance on the velocity measurements is assumed to be proportional to surface brightness, which is modeled as a S\'ersic profile of the same galaxy with $n=1$ and $R_e = 2$ kpc. The velocity inverse variance has been normalized to 0.05 (km/s)$^{-2}$, typical of a MaNGA galaxy. Right: The velocity field from the left panel but smeared and sampled coarsely to emulate an observation at $z \sim 0.7$ with a FWHM of 0.7" on a MUSE-like instrument. See details in Section \ref{sec:data}.}
	\label{samplevf}
\end{figure}

We assume that the inverse variances on the velocity measurements are proportional to surface brightness, which we set as an $n=1$ S\'ersic profile. The scale, inclination, and position angle of the S\'ersic profile are defined by the same parameters that define the mock velocity field. normalized to a peak value of 0.05 (km/s)$^{-2}$, corresponding to a velocity error of about $4.5$ km/s at the center of the galaxy. This value was chosen because it was the most common value for galaxies in MaNGA \citep{bundy15} and should capture some of the errors from approximating rotation as a thin disk as well. 

For the velocity field, we first generate an idealized toy model of a rotating galaxy using a simple model:

\begin{equation} \label{vfmodel}
v(r,\theta) = \frac{2}{\pi} v_{max} \arctan\left(\frac{r}{h}\right) \sin i\, \cos(\theta-\phi) + v_{sys}.
\end{equation}

\noindent Here $r$ and $\theta$ are polar coordinates of the spectroscopic measurements, $v_{max}$ is the asymptotic rotation speed, $h$ is the characteristic scale radius, $i$ is the inclination, $\phi$ is the kinematic PA, and $v_{sys}$ is the systemic velocity of the galaxy. The positions of the primary kinematic axes, the part of the velocity field model that the shear measurement primarily relies on, are not sensitive to the specific shape of the rotation curve model. We compared results from this model with the more complex empirically-derived Universal Rotation Curve from \cite{persic96} and found negligible difference for the results of this paper, as did \cite{wittman21} in their analysis. We then apply the transformation matrix (Equation \ref{Amatrix}) to the data to shear the velocity field, an effect more easily seen in Figure \ref{example}. For the remainder of this paper, we will use this model to both simulate mock velocity fields and to fit our mock data, allowing the model complete freedom to describe the mock galaxy.

To simulate a real ground-based observation, we first convolve the velocity field with a Moffat point-spread function (PSF) with a full-width half maximum (FWHM) of 0.7" and $\beta = 2.9$ to the velocity field and inverse variance to model atmospheric distortions, a very good night at a mountaintop observatory. We weigh the PSF convolution using the observed surface brightness profile of the galaxy. We then assume a spatial sampling of $0.2 \times 0.2$ arcsec$^2$, the same as the wide field mode for MUSE, corresponding to ${\sim} 1$ kpc spatial resolution at redshift $z = 0.7$. We generate a velocity field with a characteristic scale radius $h = 2$ kpc and apply a shear of $\gamma_{\times} \approx 0.06$ to its spatial coordinates. We assume measurements extending out to ${\sim} 2 R_e$ for an $n=1$ S\'ersic profile also with $R_e = 2$ kpc\footnote{For all other mock galaxies generated in this paper, we will always make the simplification that the characteristic scale radius of the velocity field $h$ and the effective radius of its surface brightness profile $R_e$ are equal.}, meaning we should have a grid of $8 \times 8$ spatial samples across the field of the galaxy. We also apply this same PSF convolution step when fitting this model to data in order to more accurately recover the input parameters.

We also apply random Gaussian errors in the measured velocity according to the assumed inverse variance profile, perturbing the results from the ideal velocity field. The resulting measurements, shown in Figure \ref{samplevf}, qualitatively resemble actual data taken by MUSE (see \citealt{contini16}, Fig. 5). Because we generate our mock data from ideal rotation curve models, we do not expect per-galaxy systematic errors from dynamical shape noise to be present, leaving only statistical errors. We do not apply this step when fitting the model to the data, instead using the ideal unperturbed model.

To find the best velocity field parameters for each mock galaxy, we use the Markov Chain Monte Carlo (MCMC) package \texttt{emcee} \citep{emcee}, feeding in the velocity measurements across the face of a galaxy. We use the same rotation curve model as was used to generate the velocity field, but we add shear by simply applying the inverse of the transformation matrix given in Equation \ref{Amatrix} to the underlying coordinates of the measurements and letting $\gamma_{\times}$ be a free parameter in the fit. Because our velocity field model is sheared directly by the transformation matrix, it is not affected by the approximations we made in Section \ref{sec:radon}. This, however, comes at the price of analytic simplicity of the model. 

\begin{figure*}
	\centering
	\includegraphics[width=.9\linewidth]{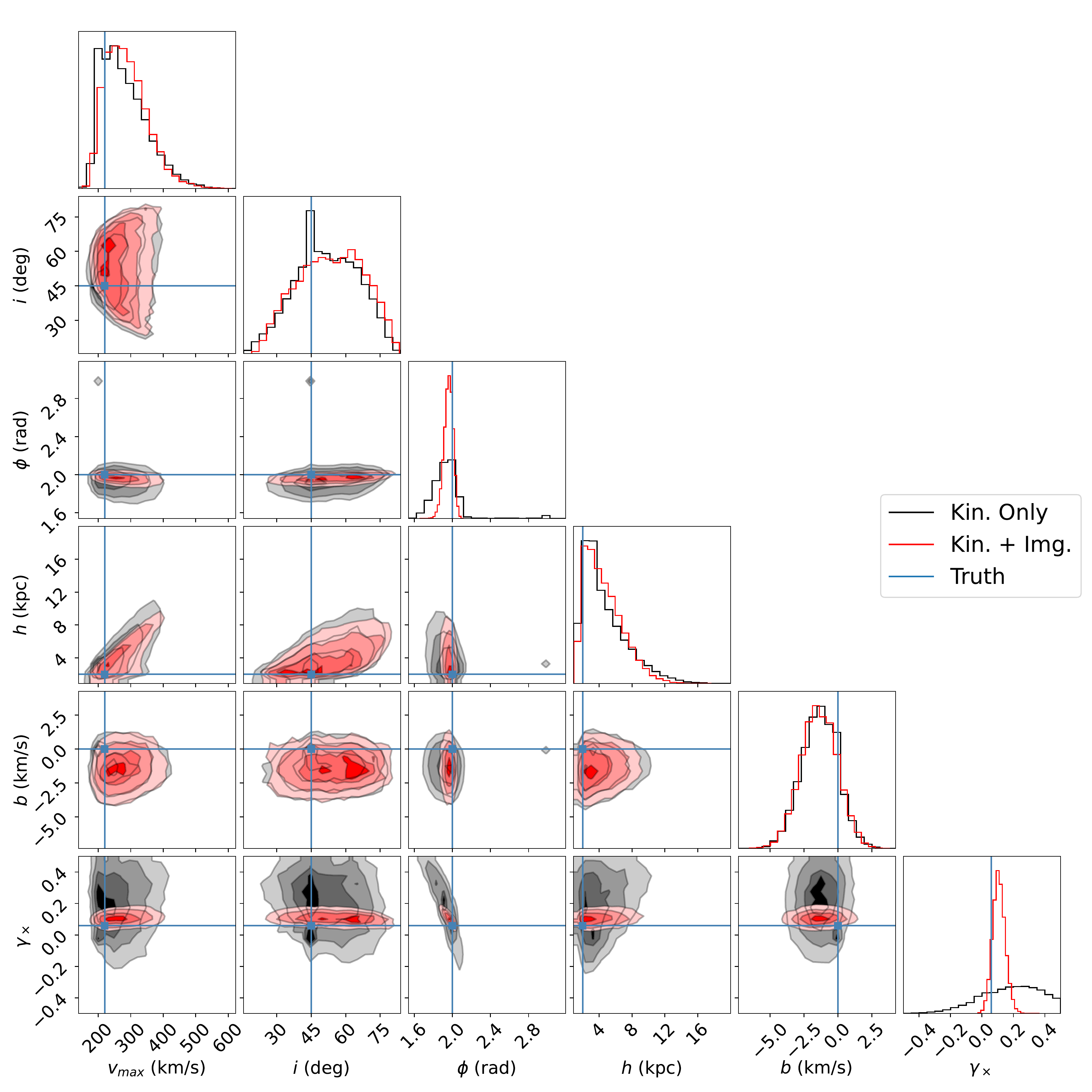}	
	\caption{Black: The resulting posteriors from applying the Bayesian velocity field model from Equation \ref{vfmodel} with a free shear parameter to the simulated sheared galaxy described in Section \ref{sec:data}. Most posteriors are centered near their true values (shown as blue lines), but the uncertainty is inflated by injected Gaussian errors, blurring from the PSF, and low spatial resolution. This is especially apparent in inclination and shear. Since the magnitude of the introduced errors greatly outweighs the signal from weak lensing, the posterior for $\gamma_{\times}$ has very poor precision and the statistical error dominates the measurement. Red: The performance of the model is greatly improved when image PA is included as described in Section \ref{sec:phot}. With a 3 degree error on the image PA, the posteriors are noticeably more constrained than the kinematics-only model, especially the shear, which has a factor of 5 reduction in spread, and the position angle $\phi$. The degeneracy between $\phi$ and $\gamma_\times$ is also resolved.}
	\label{vfcorner}
\end{figure*}

We apply uniform priors to inclination and position angle to allow for random variation in orientation. We use a Gaussian prior on asymptotic velocity with mean 200 km/s and standard deviation 100 km/s and is truncated at 0, parameters that are loosely based off of the MaNGA sample \citep{bundy15}. We allow the rotation scale to vary uniformly up to 4 arcseconds. For $\gamma_{\times}$, we apply a uniform prior bounded at $\pm 0.5$ to allow for a reasonable amount of variation, but not so much as to allow the model to fit any irregularities it sees in the data with unphysical amounts of shear. We use a standard Gaussian likelihood function for comparing the model to the mock data at each iteration. This model produces shear magnitudes and statistical errors that are similar to those measured by \cite{gurri20} when applied to the same data.

With these assumptions, we apply our Bayesian velocity-field model to the simulated data and recover the posteriors shown in black in Figure \ref{vfcorner}. We can see that the posteriors include the true values for most of the model parameters, although only some appear near the median of the posterior distributions. Two primary factors in the model precision are the effects of the PSF and the added Gaussian noise. Inclination relies on the specific shapes of isovelocity contours, but much of their variation with inclination is masked by the PSF. In addition, the width of the shear posterior $\sigma_\gamma$ is hampered by the relatively shallow velocity gradient, limited spatial resolution, and a significant degeneracy with the kinematic position angle. So while we can use this method to extract some of the parameters of the velocity field, we cannot fit our key parameter of interest with much precision. More information on the shear is needed if we want to lower statistical errors enough to produce a successful fit for data of this spatial resolution.

\section{Combining Imaging and Kinematics} \label{sec:phot}

\subsection{Kinematic and Photometric Position Angle Offset}
To better constrain the lensing distortion of the velocity field, we incorporate the image distortion we explored in Section \ref{sec:phottheory}. Comparing the kinematic axis differences derived in Equations \ref{angkin} and \ref{angkinsimp} to the photometric angular difference from Equation \ref{angphot}, we can see (as in Figure \ref{example}) that there is a difference between the angle measured by fitting an isophote and the angle from the velocity field for the major and minor axes:

\begin{multline}\label{dthmaj}
	\Delta\theta_{maj} = \gamma_\times\frac{1+q^2}{1-q^2} - \gamma_\times\\
	 = \gamma_\times\left(\frac{1+q^2}{1-q^2} - 1\right) = \frac{2 \gamma_{\times} q^2}{1-q^2}
\end{multline}

\begin{multline} \label{dthmin}
	\Delta\theta_{min} = \left(\gamma_\times\frac{1+q^2}{1-q^2} + \frac{\pi}{2} \right) - \left(\frac{\pi}{2} - \gamma_\times\right) \\
	= \gamma_\times\left(\frac{1+q^2}{1-q^2} + 1\right) = \frac{2\gamma_{\times}}{1-q^2}.
\end{multline}

\begin{figure}
\plotone{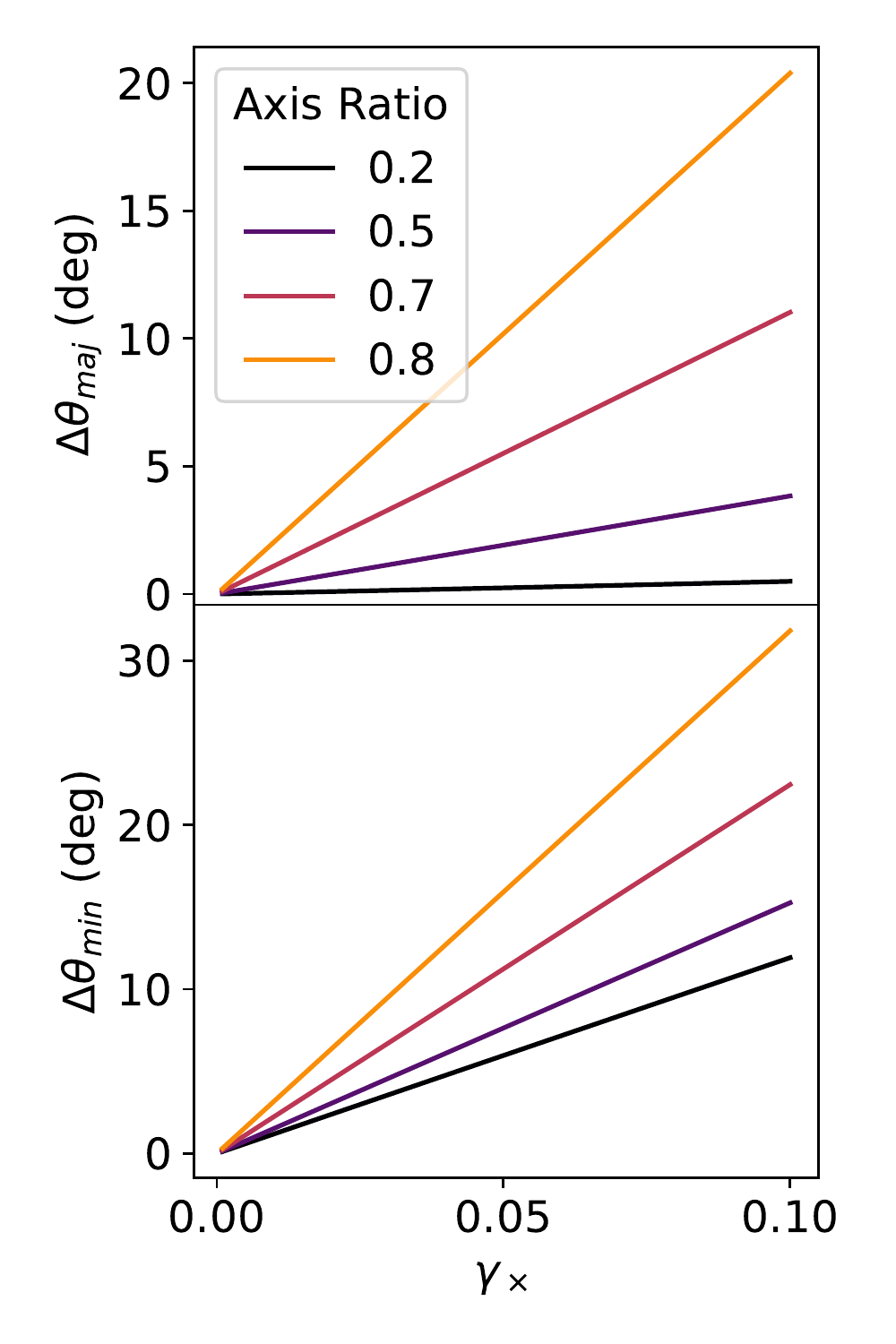}
\caption{The angular difference between the imaged and kinematic axis for the major axis (top) and minor axis (bottom) as a function of the shear $\gamma_{\times}$ and the axis ratio $q=b/a$. The magnitude of the rotation of the photometric axis increases more quickly for more face-on galaxies because less elongated ellipses are more easily distorted in an arbitrary direction than more elongated ellipses. The major axis shows slightly less angular difference than the minor axis because the imaged and kinematic axes are being shifted in the same direction, while for the minor axis, they are shifted in opposite directions. This is also why the edge-on $q=0.2$ line for the minor axis still shows some difference.}
\label{shearvsdth}
\end{figure}

\noindent The minor axis deviates more because the velocity field shearing effect goes in the opposite direction of the apparent rotation of the image. A comparison of the dependencies of $\Delta \theta$ on $\gamma_{\times}$ and $q$ for the major and minor axes is seen in Figure \ref{shearvsdth}. More face-on galaxies have larger apparent rotation angles because their shape is easy to distort. A slight elongation of an elliptical isophote in a given direction will have much more effect on the orientation of the isophote if it starts off as relatively round rather than relatively extended.

We can also see that these angular differences mainly depend on $\gamma_\times$ and $q$. We can then solve for the shear $\gamma_\times$ in terms of the two observables:  $\Delta\theta$ (the difference between measured kinematic axis and imaged axis) and the axis ratio $q$. The relationship between $\gamma_\times$ and the deflection of the major and minor axis position angles differ by a factor of $q^2$:

\begin{equation}\label{shearmaj}
	\gamma_{\times,maj} \approx \frac{\Delta\theta_{maj}(1-q^2)}{2q^2}
\end{equation}

\begin{equation}\label{shearmin}
	\gamma_{\times,min} \approx \frac{\Delta \theta_{min}(1-q^2)}{2}.
\end{equation}

\noindent The dependencies we recover here agree well with \cite{wittman21} and we will use them in the next section to derive an improved lensing inference based on a combination of imaging and kinematic measurements. By combining the imaging measurements of position angle and axis ratio with velocity measurements of the \textit{kinematic} position angle, we can gain access to a constraint on the gravitational shear induced on the galaxy.

\subsection{KWL Models with Imaging Information}

We rerun the previously described Bayesian velocity field fitting models but with added constraints on the difference between the kinematic and image PAs. We feed the model a mock image position angle, setting the value by perturbing the expected value derived using Equation \ref{dthmaj} with a Gaussian error based on the assumed level of photometric uncertainty. Commonly-used photometric codes like \textsc{Galfit} \citep{galfit} often drastically under-report the magnitude of their PA errors. Accurate error accounting must be done by comparing intrinsic PAs in simulated galaxies with the values recovered by photometric fitting. \cite{haussler07} suggests that photometric PA measurements have average errors between 1 and 6 degrees depending on image depth, so we will largely restrict our analyses to that range.

We then allow the model to fit the angular difference between the kinematic PA and the perturbed image PA with its shear parameter, incorporating the result as a new Gaussian term in the likelihood function. With this added information, we see a significant reduction in the width of the posteriors $\sigma_\gamma$, as shown by the red posteriors in Figure \ref{vfcorner}. Adding in the image PA allows for the degeneracy between the kinematic PA and the shear strength to be broken much more effectively, resulting in a factor of 5 reduction in $\sigma_\gamma$ for an image PA with an error of 3 degrees. This vast reduction in statistical errors has large implications for regimes where the measurement error would be dominated by statistical error.

\begin{figure}
	\centering
	\includegraphics[width=.9\linewidth]{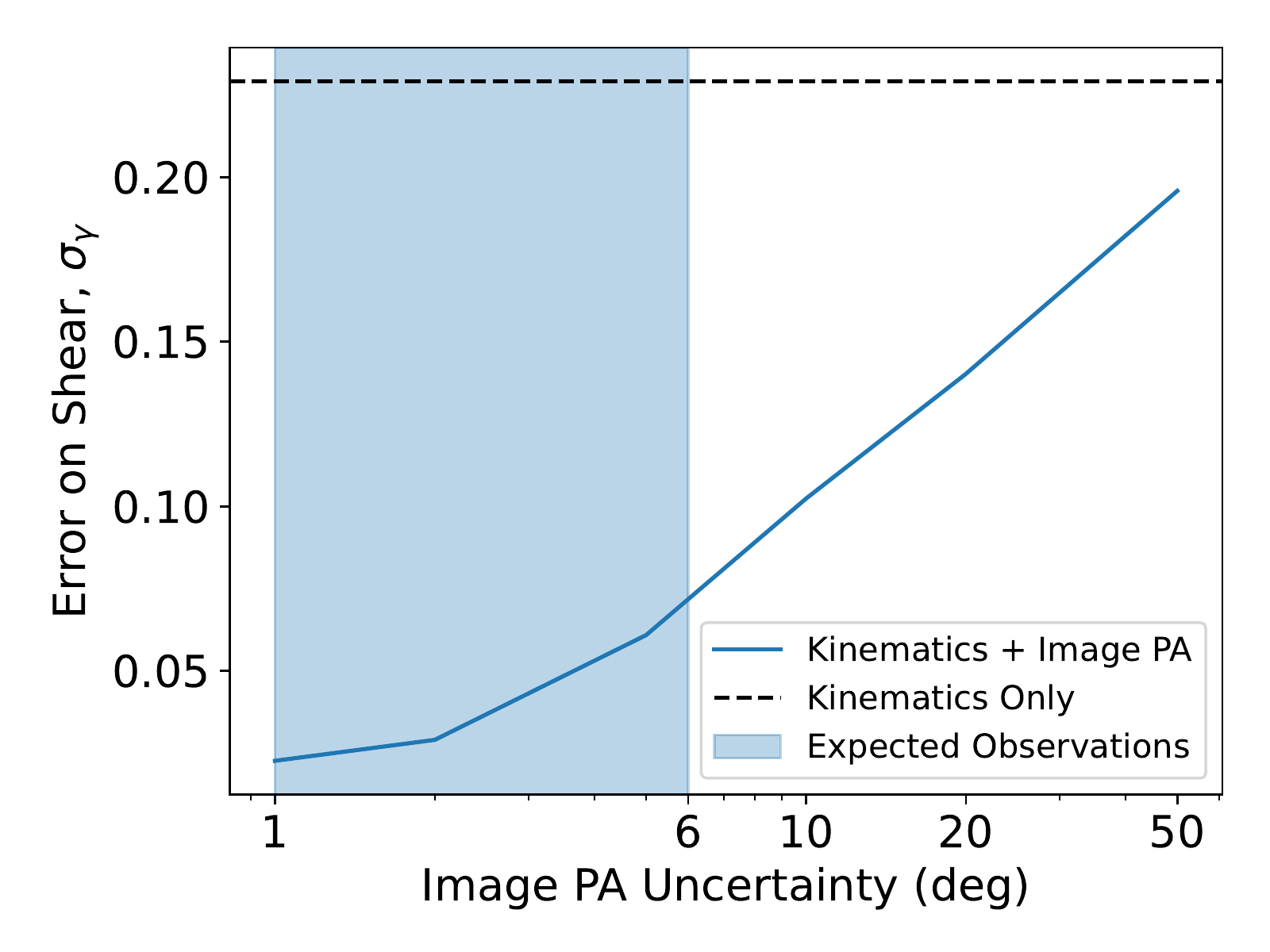}	
	\caption{
	The gains in precision seen after adding information on image position angle to our Bayesian KWL model for a mock galaxy like the one in Figure \ref{samplevf}, including the applied shear of $\gamma_\times \approx 0.06$. The model that fits both the kinematic and imaging distortion (blue line) has an error on the shear posterior $\sigma_\gamma$ that is 3-6 times smaller than the model with only kinematics (dashed line) was able to achieve within the region of expected errors on image PA in real data (blue shaded area). If borne out in real observations, this could lead to significant decreases in necessary sample sizes and exposure times to obtain a given lensing S/N. For very large errors in image PA, $\sigma_\gamma$ tends back towards the value obtained with no imaging information.}
	\label{photerrs}
\end{figure}

These improvements persist even for larger PA errors, as shown in Figure \ref{photerrs}. Within the range of expected image PA errors (shown as the shaded blue region), we see a reduction in $\sigma_\gamma$ by a factor of 2--6, meaning that a KWL technique utilizing imaging shape information is more sensitive. We also see that for large image PA errors, $\sigma_\gamma$ approaches its value from before imaging information was added. As expected, as uncertainty in the image PA measurement increases, the statistical error in the measured shear tends towards what is obtained in a fit without image PA constraints.

\begin{figure}
	\centering
	\includegraphics[width=.9\linewidth]{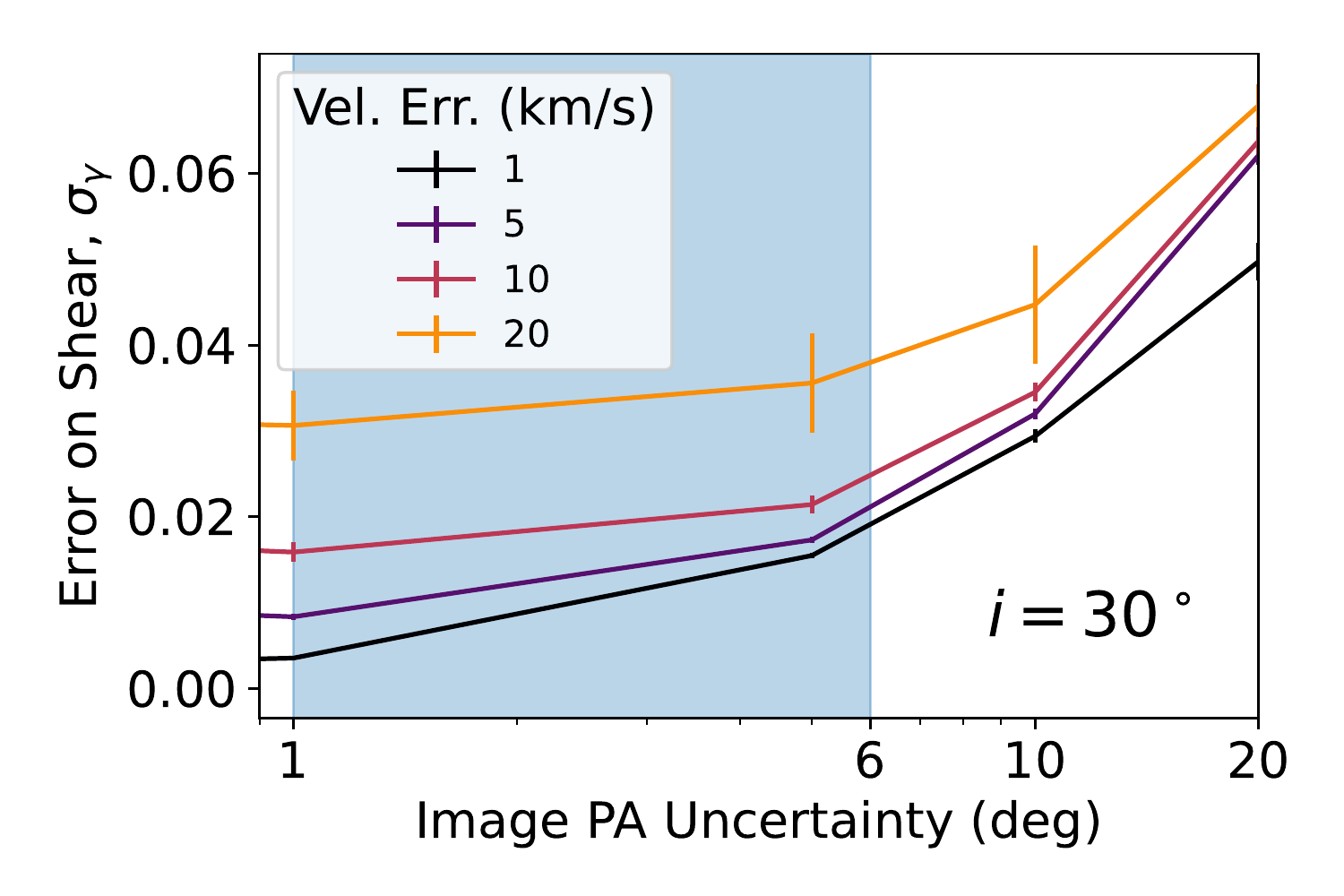}
	\caption{The change in the error on the shear measurement $\sigma_\gamma$ for a mock velocity field with simulated seeing with different amounts of error on the image PA and different central velocity errors. The input shear is $\gamma_\times \approx 0.06$ and the model velocity field is identical to those used in previous figures except that it has a $30^\circ$ inclination with respect to the observer. The blue shaded region represents the expected range for image PA errors in real observations. Smaller errors on the image PA lead to higher precision in shear, as is expected. For larger image PA errors, the precision approaches the value derived from just the velocity field, but the value it levels off at depends on the errors in the velocity field itself and the inclination of the galaxy. For image PA errors expected in real data, even poor-quality velocity fields can match the precision reached by very good velocity field-only KWL measurements, presenting opportunities for exposure time savings.}
	\label{psfgrid30}
\end{figure}

In Figure \ref{psfgrid30}, we explore differing combinations of image PA error and velocity error for a relatively face-on source galaxy (30 degree inclination). Larger errors in image PA or velocity measurement may come from shallower exposures, poor seeing or angular resolution, or irregularities in the galaxy. At this low inclination, improvements in image PA precision improve the KWL shear measurements as much or more than improvements to the velocity precision, reducing the need for higher S/N or higher resolution spectroscopy. Even with moderate image PA uncertainties on the order of several degrees, the precision of KWL measurements can reach levels comparable to those of velocity-field-only fits relying on kinematic measurements at a fraction of the velocity error.

\begin{figure}
	\centering
	\includegraphics[width=.9\linewidth]{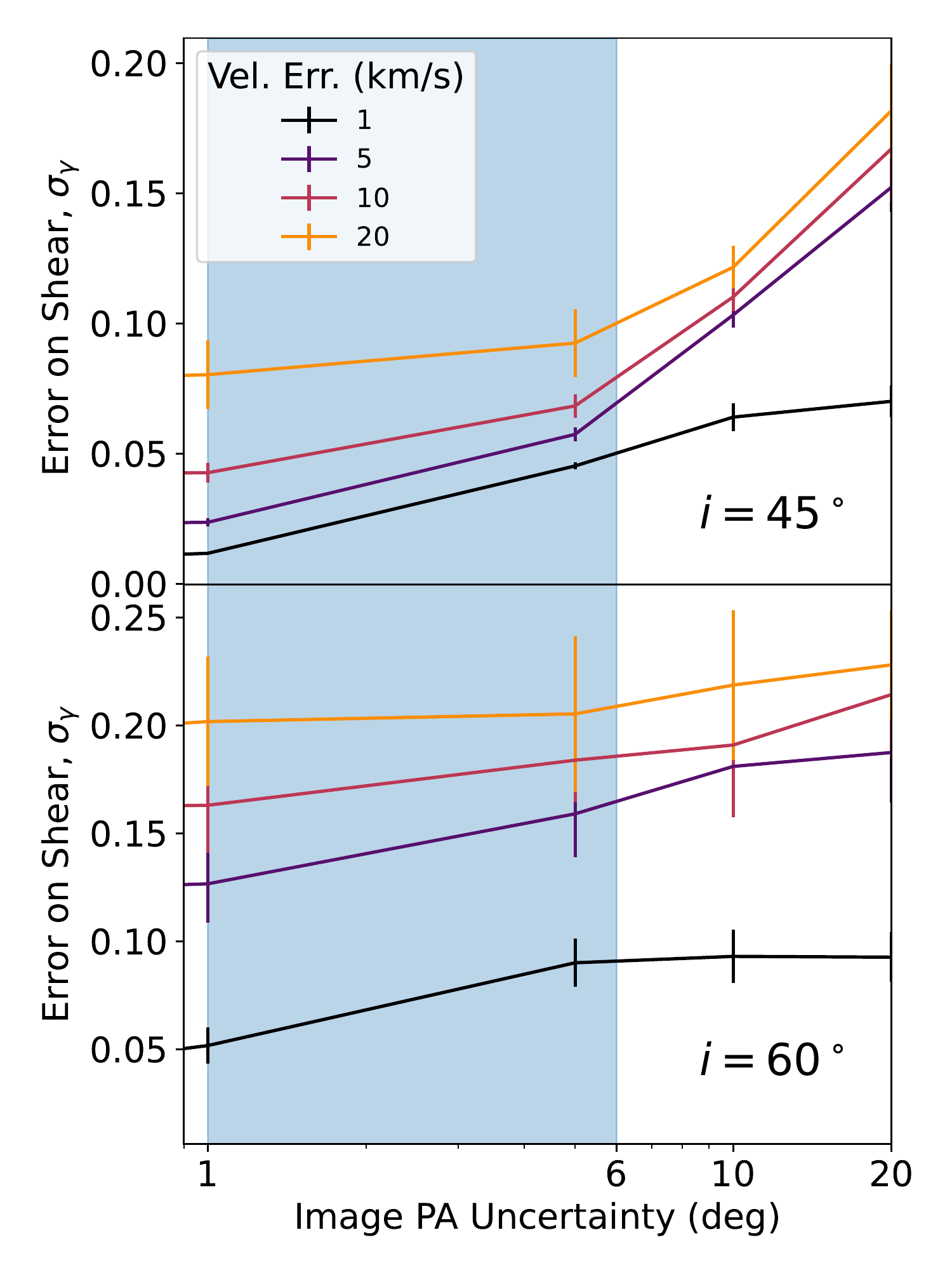}
	\caption{The same as Figure \ref{psfgrid30} but for a $45^\circ$ and $60^\circ$ galaxies. Higher-inclination galaxies have a much smaller image distortion than lower-inclination galaxies, meaning the model cannot rely on this information as much when constraining the shear. This creates a much larger stratification in velocity error than for lower-inclination galaxies. Overall, $\sigma_\gamma$ values are larger, particularly in the $60^\circ$ galaxy, but improvements to the image PA observation have little effect.}
	\label{psfgrid4560}
\end{figure}

These gains are much larger for lower-inclination galaxies since their imaged axes are much more distorted, allowing them to contribute more to the shear measurement. Higher-inclination galaxies, as shown in Figure \ref{psfgrid4560}, gain less from their photometric information because the magnitude of the angular deflection is greatly reduced, forcing the shear measurement to rely more on kinematic distortion. The distribution of inclinations within a population of randomly-oriented disks is uniform over $\sin i$ \citep{law09}, so the inclination distribution of a random sample from a survey will be weighted towards the high end, with a mean inclination angle of 1 radian (about $57^\circ$) or a mean axis ratio of $q \approx 0.54$. So unless a survey sample is deliberately designed to sample low-inclination galaxies, most galaxies will see improvements on par with Figure \ref{psfgrid4560} rather than \ref{psfgrid30}. 

Still, adding imaging information to KWL measurements for these higher-inclination galaxies that would make up the majority of a random sample would cut required exposure times dramatically for spectroscopic observations. If we assume that S/N scales with the square root of exposure time, then even a factor of two improvement in S/N from including the image PA can cut exposure time by a factor of four.

\subsubsection{Comparison to Previous Work} \label{sec:gurri}

As mentioned in Section \ref{sec:data}, our kinematics-only model delivered similar statistical errors as that of \cite{gurri20} when tested on their data, so to quantify the benefits of our kinematics + imaging technique on presently-available data sets, we again benchmark against their approach by modifying our mock data to better match the \cite{gurri20} data. We move the source galaxy to $z=0.15$ and the lens to $z=0.03$, increase the FWHM of the PSF to 1.5", and increase the spatial element size to 0.5" to roughly mimic their data set. We add mock image PA measurements, as we have done for the rest of our mock data, and we assume an error on the image PA of 3 degrees, which is in the middle of the expected error range.

Applying our kinematics + imaging algorithm to these simulated galaxies results in significantly improved statistical uncertainties.  Our estimate on the resulting $\sigma_\gamma$ values are lower by a factor of ${\sim}6$ on a per-galaxy basis, consistent with the results from higher redshift mock data in Figure \ref{photerrs}. 
This result demonstrates that even for existing data sets at lower redshifts, this technique can increase lensing precision.

However, our inclusion of imaging data does not address the dynamical shape noise term.
In order to account for this error, \cite{gurri20} estimate the magnitude of the error using a different sample of unsheared galaxies, and find an amplitude that is similar to the statistical error on a per-galaxy basis.  This dynamical shape noise term is added in quadrature to determine the final error for each galaxy in their sample.  As a result, our estimated factor of 6 improvement in the statistical error, when applied to the full sample in \citealt{gurri20}, only increases their overall 2.5$\sigma$ shear detection to 3$\sigma$.  


\subsection{Survey Design Considerations} \label{sec:survey}

While statistical error is small relative to systematic error for low redshift samples, the same is not true at higher redshifts. Resolved spectroscopic measurements become more difficult at higher redshifts due to lower spatial resolution and surface brightness, resulting in higher statistical errors. So the gains made by our imaging + kinematics model will be especially salient at increased redshifts, where there are the added benefits of greater lensing magnitudes (thanks to more favorable lensing kernels) and higher on-sky source densities, allowing for easier collection of larger sample sizes.

If we observe a $1 \times 1$ arcmin$^2$ field near a galaxy cluster (e.g. using MUSE), we will catch tens of galaxies with velocity fields that are well-defined enough to perform our KWL analysis. A multiplexed fiber IFU instrument like the proposed FOBOS \citep{fobos} would be able to patrol a significantly wider field and collect velocity data from only the most promising galaxies, enabling a larger sample in less overall exposure time. We can stack the lensing information from the multiple galaxies to obtain higher precision and reduce the random errors introduced by noise or galaxy irregularities either by a weighted average of shear results (as was done in \citealt{gurri20}) or a hierarchical Bayesian model that simultaneously fits all galaxies in a given spatial bin. We expect these individual galaxy ``dynamical shape noise" errors to average out to zero in a large sample because they will be randomly and symmetrically distributed \citep{gurri20}.

Figures \ref{psfgrid30} and \ref{psfgrid4560} show that, with attainable observational errors and low inclinations, it is possible to obtain a $\sigma_\gamma$ value comparable in magnitude to the applied shear, resulting in a ${\sim}1\sigma$ KWL measurement per galaxy for the input shear of $\gamma_\times \approx 0.06$. However, if we stack multiple such measurements within a radial bin defined by a selection of foreground lenses
and assume that the source measurements are statistically independent samples of the ensemble mass density profile of the lenses, then the error on this galaxy--galaxy shear measurement should scale as $n^{-1/2}$. So for a sample of only $n=9$ galaxies in a spatial bin, $\sigma_\gamma$ should be lowered by a factor of 3, raising a 1$\sigma$ detection to 3$\sigma$. This scaling should apply to the systematic error as well since we expect individual galaxy errors to average to zero over a large sample.

\subsubsection{Source Redshift}
Depending on the instrumentation, if we want to build a larger sample size, higher source density on-sky may be beneficial if the number of independent pointings can be minimized. Source density increases with redshift, as does the magnitude of lensing effects according to Equations \ref{sigcrit} and \ref{shear}, but there is a trade-off because surface brightness and physical size decrease with redshift. To explore this trade-off, we run our KWL model that incorporates imaging information on the same mock galaxy placed at a range of redshifts.  We ignore complications from galaxy evolution like mass/radius growth or dynamical changes. We simulate the observations of different inclination source galaxies at the varying distances as if they are behind a lens at $z=0.3$. We hold the rest of the observational parameters constant, on-sky spatial resolution, image PA error, central velocity error, and PSF width (unless specified). Maintaining constant errors would require deeper exposures for both imaging and spectroscopy, which we do not account for in this paper.

\begin{figure}
	\centering
	\includegraphics[width=.9\linewidth]{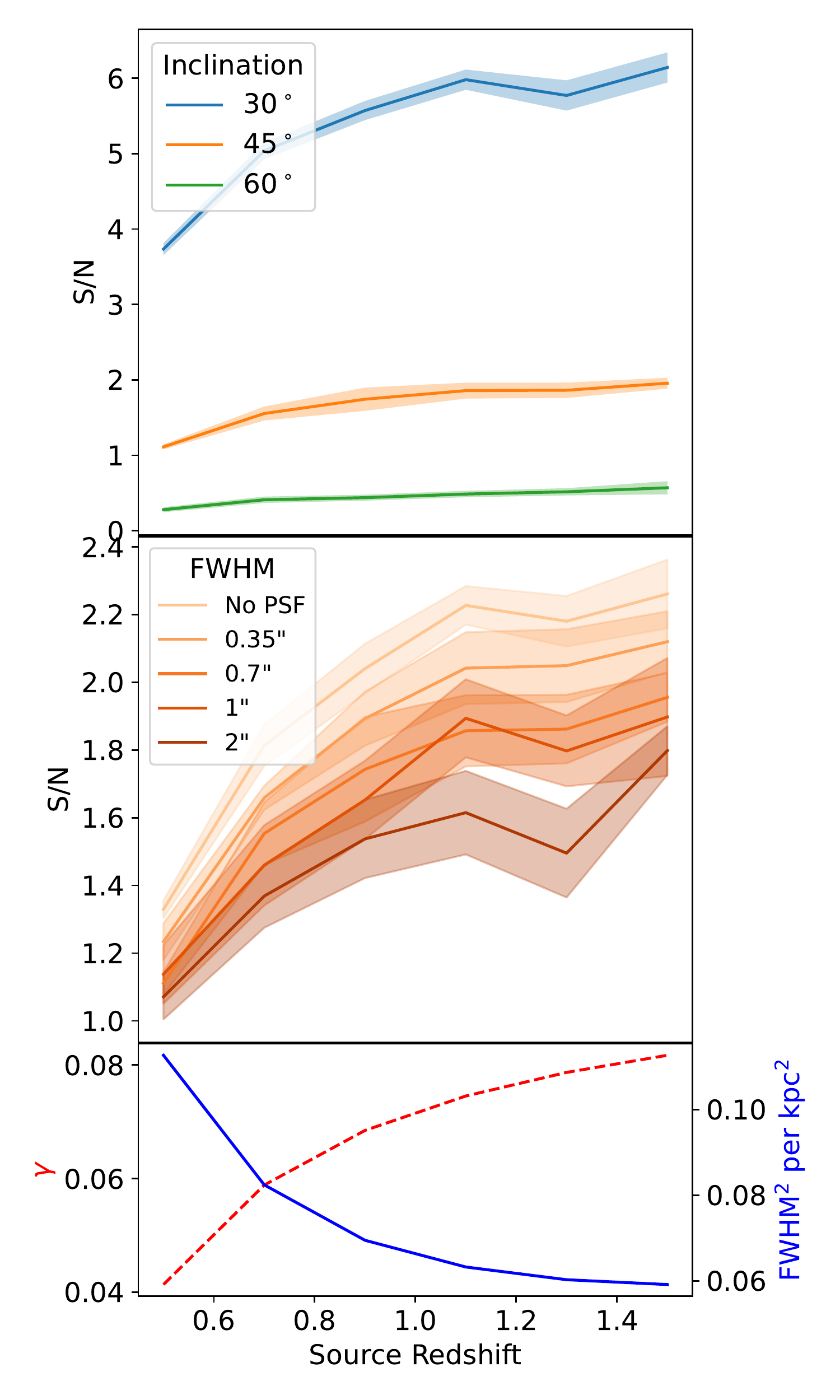}	
	\caption{Top: The KWL S/N per galaxy for source galaxies at varying inclinations with different redshifts behind a massive halo at $z=0.3$. We maintain constant on-sky spatial resolution, imaging and velocity errors, and PSF width, ignoring the effects of surface brightness dimming and galaxy evolution. Error bars are based on the random variation from running 10 identical velocity fields with distinct randomized errors. Middle: The $45^\circ$ inclination galaxy from the top panel but with FWHM of the PSF varied to simulate different observing conditions, still holding on-sky instrumental resolution constant. Realistic observing conditions provide measurements that are a factor of 2--3 lower in S/N compared to an ideal observation without PSF smearing. Improving the PSF by a factor of 2 with something like a ground-layer adaptive optics system to 0.35" leads to S/N gains of about 50\%. Bottom: The overall magnitude of the lensing shear, shown in the red dashed line, increases with redshift, making the effects of KWL more noticeable. The fraction of a square FWHM covered by a square kpc in the on-sky plane of the source galaxy, shown in blue solid line, indicates that the PSF becomes very large compared to the spatial scales of the source galaxy, worsening the quality of the fit. S/N largely tracks with shear magnitude.}
	\label{snz_zl3}
\end{figure}

Although closer source galaxies have smaller $\sigma_\gamma$ uncertainties because of their greater number of spatial resolution elements, Figure \ref{snz_zl3} shows that they deliver a poorer S/N per galaxy because they are simply not sheared as much as more distant galaxies.
The top panel shows a high degree of S/N stratification in inclination because the change in image PA is much greater for lower inclinations. We find that there is modest gain in S/N for all inclinations with redshift before leveling off around $z\sim0.7-1$, the redshift range assumed in this paper, before keeping a relatively constant value at higher $z$. This mirrors the trend seen in the overall shear magnitude seen as the red dashed line in the bottom panel of Figure \ref{snz_zl3}.

Another factor that influences precision is the FWHM of the spatial PSF of the resolved kinematic field. The blue line in the bottom panel of \ref{snz_zl3} shows that as the source galaxy redshift increases, the size of the galaxy on-sky becomes smaller relative to the size of the PSF, here represented as the ratio of the square FWHM compared to the size of a square kpc in the plane of the galaxy. 
Reductions to the FWHM result in higher surface brightness in many cases and more independence of the spatial resolution elements of an IFU observation, which would improve the precision of the velocity field fit even if the spatial sampling is not changed.
These types of gains could be obtained in the real world using a ground-layer adaptive optics system \citep[e.g.][]{chun18, hartke20}, which would lead to an improvement in seeing by a factor of 1.5 to 2 over a several arcminute-wide field. In the middle panel of Figure \ref{snz_zl3}, we vary the FWHM to simulate different observing conditions, from an ideal observation without seeing to a FWHM of 2". We find modest gains in S/N with better seeing, especially at moderate and higher redshifts. However, even fits with poor seeing are able to locate the kinematic axes with acceptable precision, so the effects of better seeing are limited. Still, the increase in surface brightness is valuable for driving down exposure time. 

\subsubsection{Lens Redshift}

\begin{figure}
	\centering
	\includegraphics[width=.9\linewidth]{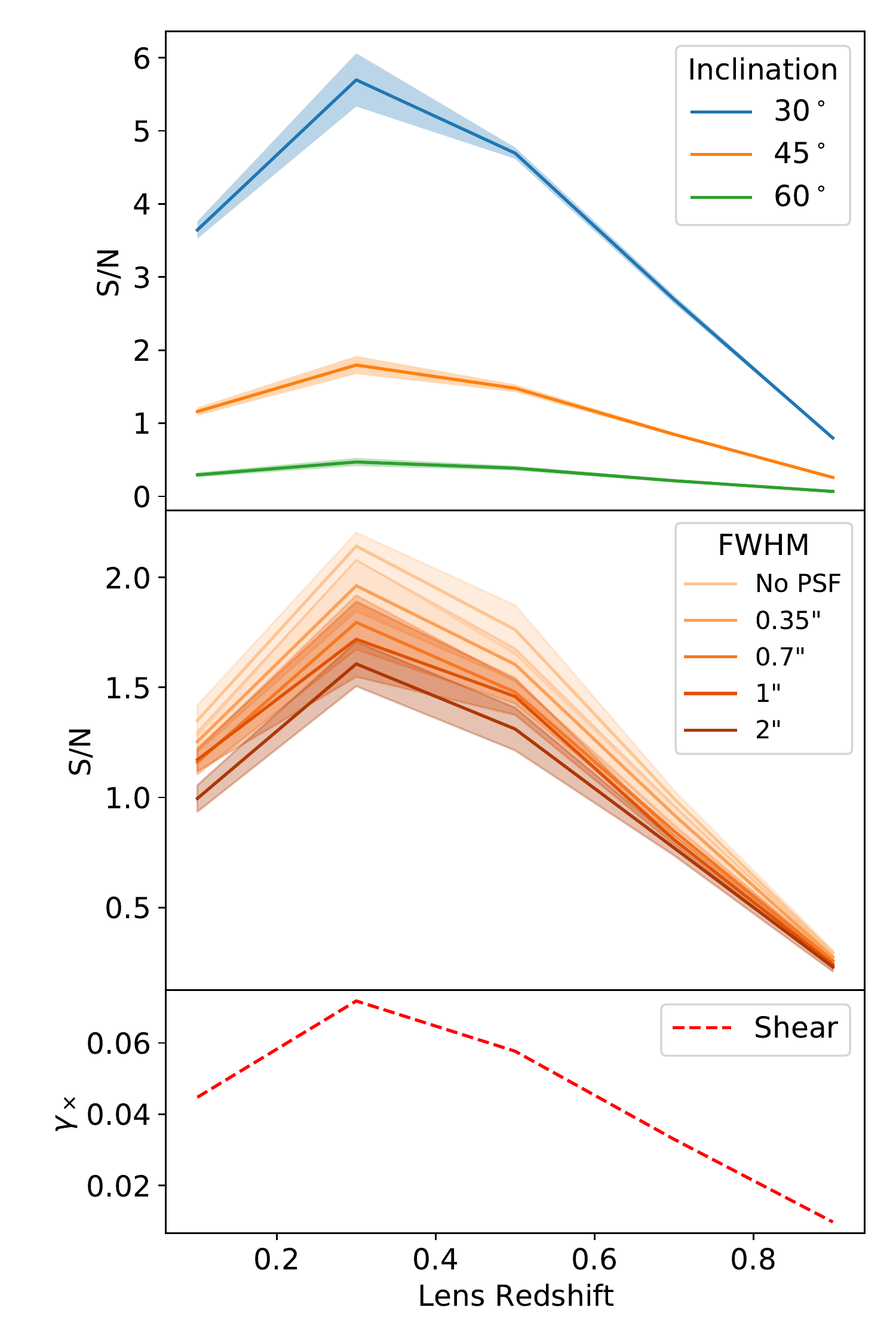}	
	\caption{The same as Figure \ref{snz_zl3} but holding the source galaxy fixed at $z=1$ while moving the lens in front of it. The S/N again largely follows the shear magnitude (bottom), which peaks at $z \sim 0.25-0.3$ because that is the configuration of angular diameter distances that minimizes the critical radius (Equation \ref{sigcrit}). FWHM$^2$ per kpc$^2$ is not shown because it is constant due to the stationary source galaxy.}
	\label{snz_zs1}
\end{figure}

To test KWL's ability to probe lenses at various redshifts, we also simulate systems where the source is held at a constant $z=1$ for varying redshift lenses (Figure \ref{snz_zs1}). We find similar dependencies on inclination and FWHM as in Figure \ref{snz_zl3}. However, since shear magnitude scales with the angular diameter distances to the source and the lens (Equation \ref{sigcrit}), shear magnitude peaks at $z\sim0.3$, falling off sharply from there. In fact, due to how angular diameter distance is defined, for source galaxies further than $z\sim0.7$, shear magnitude will always be largest for lenses at $0.2 < z < 0.3$. So KWL, like any other lensing technique, will be most sensitive to lenses at these moderate redshifts.

\subsubsection{General Strategy}
An ideal strategy for a KWL survey would likely be to target low-inclination galaxies that benefit more from the gains imaging can provide. They should be regularly-rotating blue spiral galaxies that are likely to have a high emission line flux for ease of kinematic measurements.  
The targets should have regular, symmetric isophotes and imaging sufficient to measure image position angles with 1--3 degree uncertainties. 
If the source galaxy population reaches $z \sim 1$, the ideal lens sample would have a redshift range $z = 0.2$--0.7 to ensure shear magnitudes near the peak of the lensing kernel. 
Ideally, the major-axis position angles of selected source galaxies should be misaligned with respect to the on-sky direction to the lens center.  A $45^\circ$ offset maximizes the cross-term in the shear.  
A full optimization of a KWL survey design would take into account the lens and source redshift distributions, the intrinsic variation in galaxy shape, systematic errors in individual galaxies due to dynamical shape noise, and observational constraints such as seeing and exposure time as it relates to brightness dimming, distance, and size.

\section{Conclusion} \label{sec:conclusions}

We have demonstrated a new formalism for combining imaging and kinematic information that significantly improves the S/N of kinematic weak-lensing observations. Our Bayesian models fit both the kinematic distortion and the photometric offset caused by lensing shear in a given galaxy, leading to decreases in per-galaxy statistical error by a factor of 2 to 6 compared to kinematics alone.
If borne out in real observations, this approach opens the door for future studies to more effectively utilize kinematics-based lensing observations of lens samples at moderate redshifts.
Even with existing instruments, the methodology appears promising for enabling probes of galaxy cluster halo profiles with greater spatial resolution and S/N 
while mitigating many of the systematics that affect conventional lensing measurements, like shape noise and intrinsic alignments. Ultimately, this will allow 
for individual, total mass measurements in the weak-lensing regime for a greater number of clusters. 

A purpose-designed survey on a new or existing instrument could provide the observations necessary for a successful KWL measurement. Further study is needed to determine the appropriate redshift range to balance shear magnitude and source density with spatial resolution and surface brightness, while more detailed galaxy rotation models are needed to quantify the systematic effects of kinematic irregularities on KWL measurements.

\acknowledgments
The authors would like to acknowledge the anonymous reviewer for their helpful comments in improving this paper. We woudl also like to acknowledge Eric Huff and Tim Eifler for their early guidance on the theory of kinematic weak lensing and Kevin McKinnon for early statistical guidance. 

\bibliography{kwldetection}



\end{document}